\documentclass[reprint,prl,aps]{revtex4-1}
\usepackage{graphicx}
\usepackage{color}
\usepackage{dcolumn}
\usepackage{bm}
\usepackage{amssymb}
\usepackage{color,amsmath}
\usepackage{soul,xcolor} 
\setstcolor{red} 
\usepackage{epstopdf}

\newcolumntype{C}[1]{>{\centering\arraybackslash}m{#1}}
\newcolumntype{N}{@{}m{0pt}@{}}

\definecolor{cadmiumgreen}{rgb}{0.0, 0.42, 0.24}

\begin{document}

\title{Anomalous Hall effect at half filling in twisted bilayer graphene}

\author{Chun-Chih Tseng$^{*1}$}
\author{Xuetao Ma$^{*2}$}
\author{Zhaoyu Liu$^{*1}$}
\author{K. Watanabe$^{3}$}
\author{T. Taniguchi$^{4}$}
\author{Jiun-Haw Chu$^{1}$}
\author{Matthew Yankowitz$^{1,2\dagger}$}

\affiliation{$^{1}$Department of Physics, University of Washington, Seattle, Washington, 98195, USA}
\affiliation{$^{2}$Department of Materials Science and Engineering, University of Washington, Seattle, Washington, 98195, USA}
\affiliation{$^{3}$Research Center for Functional Materials, National Institute for Materials Science, 1-1 Namiki, Tsukuba 305-0044, Japan}
\affiliation{$^{4}$International Center for Materials Nanoarchitectonics, National Institute for Materials Science,  1-1 Namiki, Tsukuba 305-0044, Japan}
\affiliation{$^{*}$These authors contributed equally to this work.}
\affiliation{$^{\dagger}$myank@uw.edu (M.Y.)}

\maketitle

\textbf{Magic-angle twisted bilayer graphene (tBLG) has been studied extensively owing to its wealth of symmetry-broken phases, correlated Chern insulators, orbital magnetism, and superconductivity~\cite{Cao2018a,Cao2018b,Yankowitz2019,Lu2019,Sharpe2019,Serlin2020,Balents2020,Andrei2020}. In particular, the anomalous Hall effect (AHE) has been observed at odd integer filling factors ($\nu=1$ and $3$) in a small number of tBLG devices~\cite{Sharpe2019,Serlin2020,Stepanov2021}, indicating the emergence of a zero-field orbital magnetic state with spontaneously broken time-reversal symmetry~\cite{Zhang2019,Zhang2019ahe,Bultinck2020ahe}. However, the AHE is typically not anticipated at half filling ($\nu=2$) owing to competing intervalley coherent states~\cite{Zhang2019,Bultinck2020kivc,Lian2021,Kwan2021iks,Wagner2021iks}, as well as spin-polarized and valley Hall states that are favored by an intervalley Hund's coupling~\cite{Zhang2019,Khalaf2020,Wagner2021iks,Lin2022}. Here, we present measurements of two tBLG devices with twist angles slightly away from the magic angle (0.96$^{\circ}$ and 1.20$^{\circ}$), in which we report the surprising observation of the AHE at $\nu=+2$ and $-2$, respectively. These findings imply that a valley-polarized phase can become the ground state at half filling in tBLG rotated slightly away from the magic angle. Our results reveal the emergence of an unexpected ground state in the intermediately-coupled regime ($U/W \sim 1$, where $U$ is the strength of Coulomb repulsion and $W$ is the bandwidth), in between the strongly-correlated insulator and weakly-correlated metal, highlighting the need to develop a more complete understanding of tBLG away from the strongly-coupled limit.}

Twisted bilayer graphene features eight degenerate flat bands arising from its two-fold spin, valley, and sublattice degrees of freedom~\cite{Bistritzer2011,Bultinck2020kivc}. Around the magic angle of $\theta \sim 1.1^{\circ}$, strong Coulomb interactions can spontaneously lift these degeneracies, driving emergent correlated ground states tuned by the charge doping~\cite{Balents2020,Andrei2020}. The eight-fold degeneracy of the flat bands enables a wealth of competing symmetry-broken ground state orders, including states with complete spin and/or valley polarization~\cite{Zhang2019,Zhang2019ahe,Zhang2020ivc,Bultinck2020ahe,Liu2021qah} and various forms of partially or fully intervalley coherent states~\cite{Po2018,Zhang2019,Zhang2020ivc,Bultinck2020kivc,Lian2021,Kwan2021iks,Wagner2021iks}. Although a complete understanding of the correlated ground states and their precise connection to superconductivity remains lacking, calculations performed in the strong-coupling limit of $U/W >> 1$ have so far predicted ground state orders that are broadly consistent with experimental observations at each integer $\nu$.

\begin{figure*}[t]
\includegraphics[width=6.9 in]{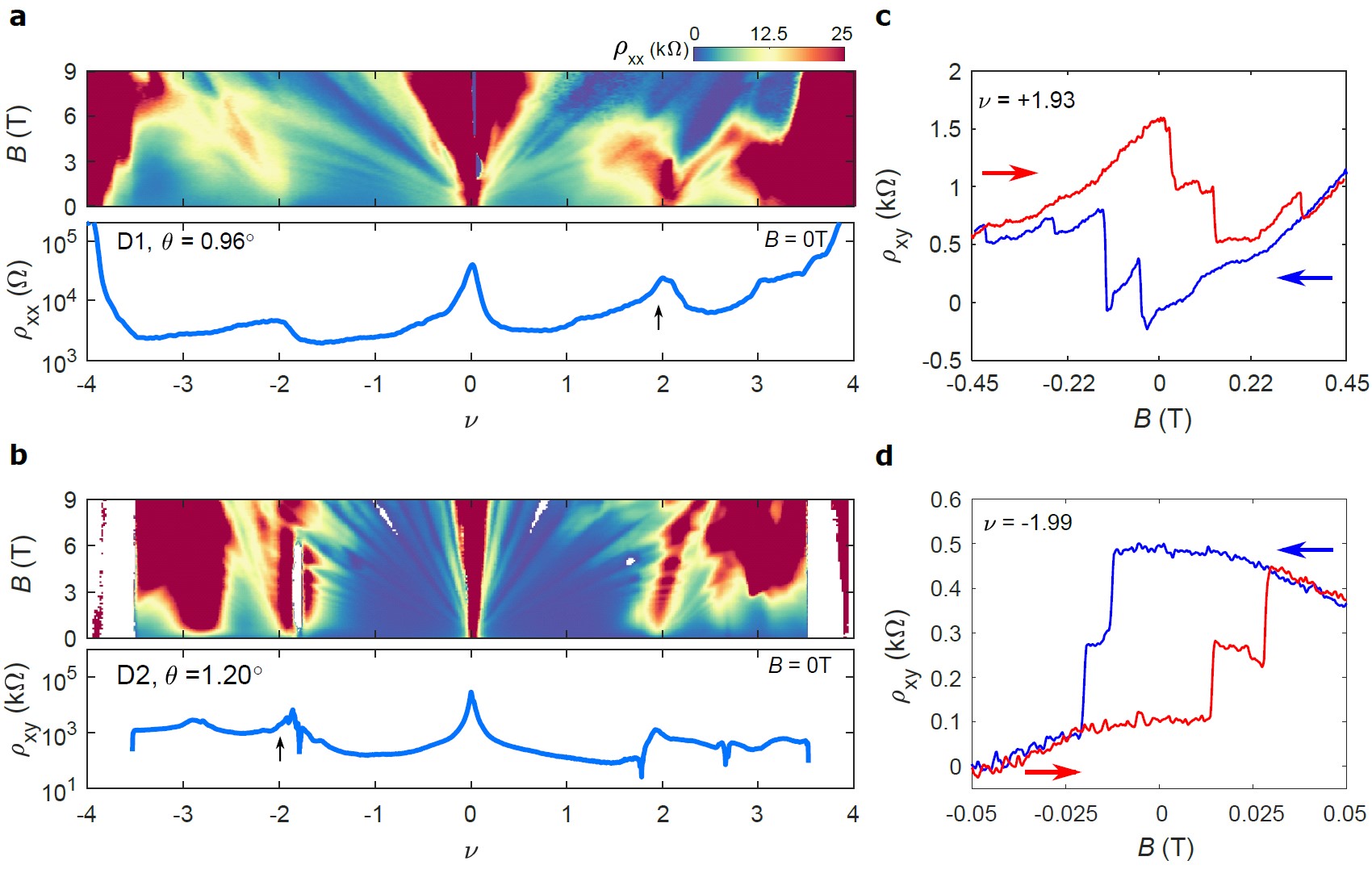} 
\caption{\textbf{Anomalous Hall effect at half filling in tBLG.}
\textbf{a-b}, (top) Landau fan diagram of the longitudinal resistivity, $\rho_{xx}$, for Device D1 ($\theta=0.96^{\circ}$) and D2 ($\theta=1.20^{\circ}$), respectively. (bottom) $\rho_{xx}$ measured at $B=0$. Regions of white in the Landau fan in \textbf{(b)} correspond to artifacts of negative measured resistance. Similar artifacts manifest as abrupt dips in $\rho_{xx}$ at $B=0$ (see Methods for further discussion of these features).
\textbf{c}, Hall resistance, $\rho_{xy}$, in Device D1 measured as the magnetic field is swept back and forth at $\nu=+1.93$ (position of the black arrow in \textbf{a}). 
\textbf{d}, Similar measurement of $\rho_{xy}$ in Device D2 at $\nu=-1.99$ (position of the black arrow in \textbf{b}).
The data for Device D1 (D2) is acquired at a nominal base temperature of 20 (100)~mK. 
}
\label{fig:1}
\end{figure*}

The ratio of $U/W$ describing the interaction strength decreases as the twist angle is tuned away from the magic angle, and eventually the band dispersion can no longer be treated as a small perturbation to the Coulomb energy. This regime of intermediate-coupling strength, in which $U/W \sim 1$, has been the subject of substantially less theoretical and experimental attention so far, but may hold important clues necessary for developing a complete understanding of the many-body phase diagram of tBLG. Here, we report transport measurements of devices at twist angles slightly larger than and smaller than the magic angle. We observe an unexpected anomalous Hall effect at half filling in these devices, arising from a new correlated ground state with finite orbital magnetization that is most naturally explained by the formation of a valley-polarized phase. However, valley-polarized states are generally not predicted to be favored in the strong-coupling limit at half-filling~\cite{Zhang2019,Bultinck2020kivc,Lian2021,Kwan2021iks,Wagner2021iks}, nor have they been observed at zero field in any previously reported tBLG device (except with the addition of proximity-induced spin-orbit coupling~\cite{Lin2022}). This suggests that competing symmetry-broken orders can become preferred as the interaction strength is reduced by tuning away from the magic angle, necessitating a more careful theoretical treatment of tBLG in the intermediately-coupled regime.

\begin{figure}[t]
\includegraphics[width=3 in]{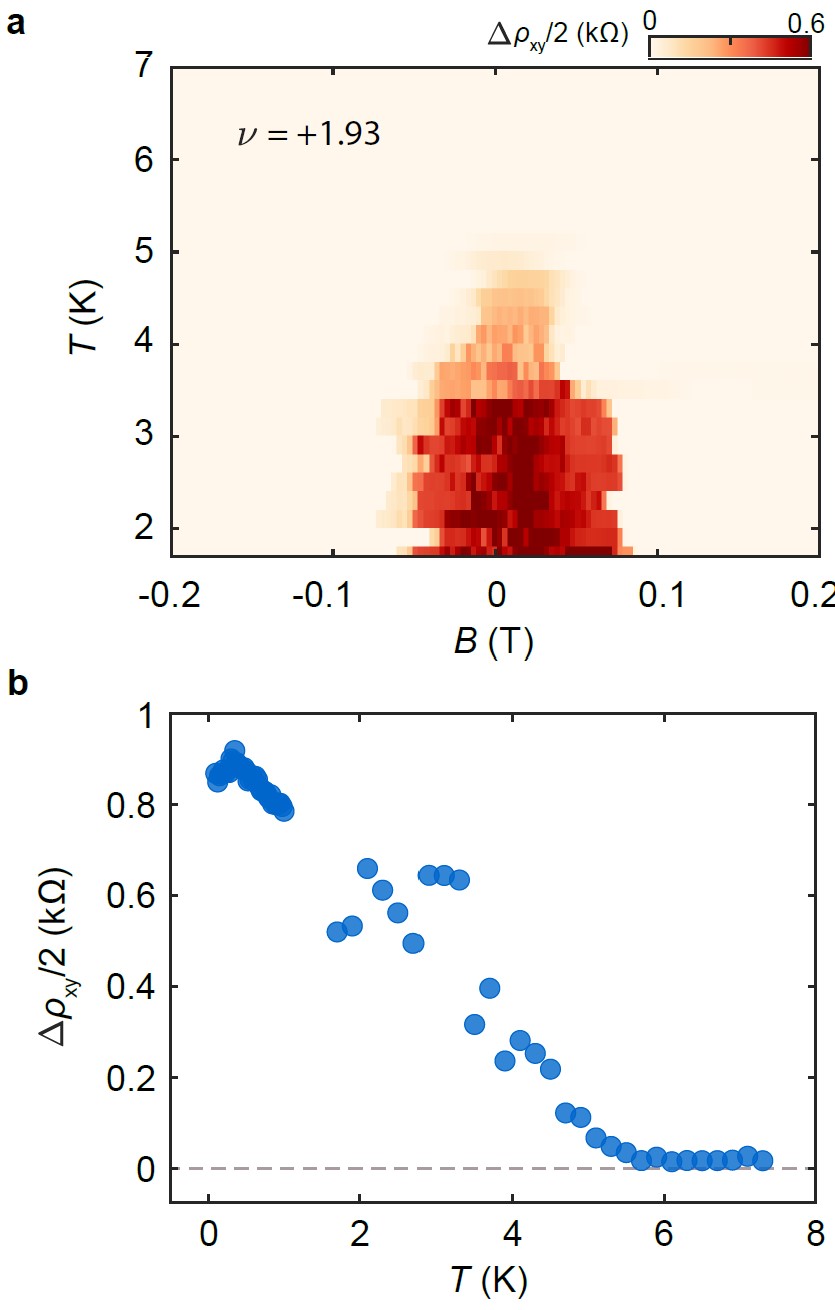} 
\caption{\textbf{Temperature dependence of the AHE in Device D1.}
\textbf{a}, AHE amplitude, $\Delta \rho_{xy}/2 = (\rho_{xy}^{B\downarrow}-\rho_{xy}^{B\uparrow}$)/2, measured at $\nu=+1.93$ as $B$ is swept back and forth at various temperatures.
\textbf{b}, $\Delta \rho_{xy}/2$ at $B=0$ versus temperature.
}
\label{fig:2}
\end{figure}

Figures~\ref{fig:1}a-b show Landau fan diagrams for our two devices (D1 with $\theta=0.96^{\circ}$ and D2 with $\theta=1.20^{\circ}$, respectively). Below each, we plot the device resistivity within the flat band ($-4 \leq \nu \leq +4$) acquired in the absence of a magnetic field and at our nominal base temperature of $T \leq 100$~mK. We observe many features typical of magic-angle tBLG at high field, including a cascade of correlated Chern insulator states and associated fans of quantum oscillations projecting to various integer $\nu$~\cite{Cao2018a,Cao2018b,Yankowitz2019,Lu2019,Wong2019,Zondiner2019}. However, we do not observe robust zero-field insulating states for any integer $\nu$, most likely indicating that the samples are in a moderately correlated regime.

Our primary result is the observation of the AHE within a small region of doping around half filling. Figure~\ref{fig:1}c (d) shows a representative measurement of the Hall resistance, $\rho_{xy}$, in Device D1 (D2) as the magnetic field, $B$, is swept back and forth at $\nu=+1.93$ ($-1.99$). We observe hysteresis and Barkhausen jumps indicating the presence of magnetism in these states. Figure~\ref{fig:2}a shows the temperature dependence of the AHE in Device D1 at $\nu=+1.93$, acquired by taking the average difference of $\rho_{yx}$ between the two field sweeping directions, $\Delta \rho_{xy}/2 = (\rho_{xy}^{B\downarrow}-\rho_{xy}^{B\uparrow}$)/2. Figure~\ref{fig:2}b plots $\Delta \rho_{xy}/2$ at $B=0$ as a function of temperature. Although the Curie temperature is $\sim5.5$~K, the amplitude of the AHE grows slowly as the device is cooled to base temperature, and saturates to a value of less than 1~k$\Omega$. Taken at face value, these observations imply that the ground state at $\nu=+2$ is an ungapped symmetry-broken metal, since the AHE amplitude is far from the quantized value of $h/Ce^{2}$ for any reasonable integer value of the Chern number, $C$. However, twist-angle disorder arising due to unintentional strain in the device may instead play a role in obscuring a small intrinsic gap at $\nu=+2$. In either case, our observations sharply contrast the behavior of the trivial correlated insulating states that are conventionally observed in devices closer to the magic angle, which exhibit insulating longitudinal and Hall resistance without associated hysteresis~\cite{Cao2018a,Cao2018b,Yankowitz2019,Lu2019}.

\begin{figure*}[t]
\includegraphics[width=6.9 in]{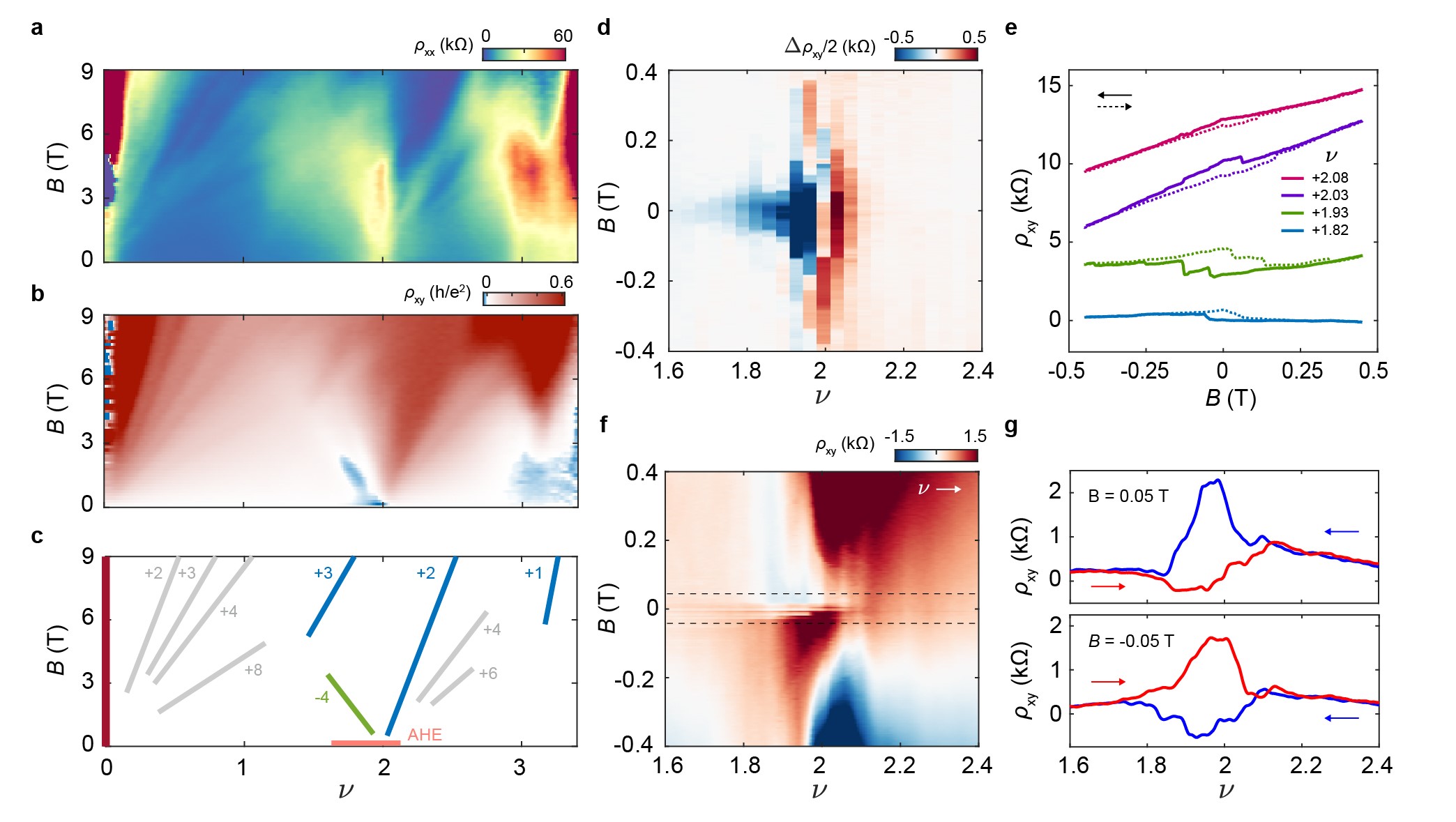} 
\caption{\textbf{Symmetry-breaking at half filling and electrical switching of the magnetic state.}
\textbf{a}, Field-symmetrized Landau fan diagram of the longitudinal resistivity, $\rho_{xx}$, of Device D1.
\textbf{b}, Field-antisymmetrized Landau fan diagram of the Hall resistance, $\rho_{xy}$.
\textbf{c}, Schematic diagram denoting the most robust gapped states observed in the Landau fan. Trivial insulators are denoted in red, correlated Chern insulators in blue, and additional quantum oscillations in gray and green. The horizontal orange line denotes the range of doping over with the AHE is observed.
\textbf{d}, $\Delta \rho_{xy}/2$ acquired by sweeping $B$ back and forth at fixed $\nu$.
\textbf{e}, Selected $\rho_{xy}$ traces from (\textbf{d}). Curves are offset vertically for clarity.
\textbf{f}, $\rho_{xy}$ acquired by sweeping $\nu$ from small to large values at fixed $B$.
\textbf{g}, $\rho_{xy}$ traces at $B=\pm50$~mT acquired for both sweeping directions of $\nu$, as indicated by the gray dashed lines in (\textbf{f}).
The data is acquired at $T= 20$~mK in all measurements.
}
\label{fig:3}
\end{figure*}

Figure~\ref{fig:3}a shows a zoomed-in field-symmetrized Landau fan diagram of $\rho_{xx}$ for Device D1. At high field, we observe a fan of two-fold degenerate quantum oscillations that project to $\nu=+2$ at $B=0$ and disperse towards larger filling factor, away from the charge neutrality point (CNP, $\nu=0$). These indicate the formation of a new Fermi surface with reduced size compared with the isospin-unpolarized phase nearer the CNP, as has been observed regularly in magic-angle tBLG owing to a doping-dependent symmetry-breaking cascade~\cite{Wong2019,Zondiner2019}. Figure~\ref{fig:3}b shows the corresponding field-antisymmetrized measurement of the Hall resistance, $\rho_{xy}$. At slight underdoping of $\nu=+2$, we see a region in which the Hall effect reverses sign, corresponding to a weakly-developed quantum oscillation with an apparent Chern number of $-4$ projecting towards smaller filling factor. This state is eventually interrupted at high field by the formation of a $C=+3$ correlated Chern insulator projecting to $\nu=1$ at $B=0$. Figure~\ref{fig:3}c summarizes the most robust gapped states we observe, with trivial insulators denoted in red, correlated Chern insulators in blue, and additional quantum oscillations in gray and green. The $\rho_{xy}$ sign reversal and the associated $C=-4$ state depicted by the green line in Fig.~\ref{fig:3}c contrast the typical behavior of magic-angle tBLG devices, in which the cascade of symmetry-breaking transitions arise only very near each integer filling and result in fans of quantum oscillations that disperse exclusively towards larger band filling~\cite{Cao2018a,Cao2018b,Yankowitz2019,Lu2019}. The precise details of these isospin Stoner transitions depend sensitively on the value of $U/W$~\cite{Zondiner2019}, potentially accounting for the difference in the behavior of the quantum oscillations we observe in this device.

Although the $\rho_{xy}$ sign reversal persists to $B=0$, careful measurements of the AHE suggest that its origin is different below and above the coercive field, $B_c$. For $B>B_c$, the sign change arises owing to the $C=-4$ state discussed earlier. For $B<B_c$, the sign change instead arises as a consequence of a reversal of the orbital magnetic state upon doping. We perform two distinct measurements in order to clearly identify the latter effect. First, in Fig.~\ref{fig:3}d we plot $\Delta \rho_{xy}/2$ acquired by sweeping $B$ back and forth at fixed $\nu$, and observe an abrupt sign change precisely at $\nu=+2$. Figure~\ref{fig:3}e shows $\rho_{xy}$ traces at selected values of $\nu$, in which we see that the sense of the AHE flips upon doping across half filling. Second, in Fig.~\ref{fig:3}f we plot $\rho_{xy}$ acquired by sweeping $\nu$ from positive to negative at fixed $B$ (Supplementary Fig.~\ref{fig:S_dopingAHE} shows a comparable measurement for the opposite sweeping direction). Figure~\ref{fig:3}g shows corresponding traces at $B=\pm50$~mT acquired for both sweeping directions of $\nu$. For a given sign of $B$, the magnetic state is determined purely by the gate sweeping direction.

Similar electric-field--reversal effects have been reported previously at $\nu=+1$ and $+3$ in tBLG~\cite{Polshyn2020,Grover2022} and at $\nu=+3$ in twisted monolayer-bilayer graphene~\cite{Polshyn2020}, all of which exhibit a nearly-quantized AHE. Such an effect can arise from the interplay between the bulk orbital magnetic moment and that of the chiral edge mode~\cite{Zhu2020orbital,Polshyn2020,Grover2022}. If the sign of the magnetization (i.e., the sum of the bulk and edge contributions) changes as the chemical potential crosses integer $\nu$, it can become favorable for the system to undergo a first-order phase transition in which carriers recondense into the opposite valley in order to maintain the alignment of the total magnetization with the external field. Similar effects are also feasible even if the state is not completely gapped, as may be the case in our sample.

Prior measurements of the magnetism underlying the AHE in tBLG indicate that it is driven primarily or exclusively by orbital magnetic moments~\cite{Tschirhart2021,Sharpe2021,Grover2022}. Owing to the extraordinarily weak spin-orbit coupling (SOC) in graphene, the AHE we observe here is almost certainly also driven by orbital magnetism. Consequently, a correlated ground state at half filling with a spontaneous valley imbalance is the most natural mechanism consistent with all of our observations described above. Such a state spontaneously breaks time-reversal symmetry, resulting in an AHE due to the large Berry curvature concentrated at the band extrema. In $C_{2}T$-symmetric tBLG ($C_2$ is a two-fold rotation and $T$ is time reversal), Hartree-Fock calculations performed in the strong-coupling limit predict either a Kramers intervalley coherent (K-IVC) or an incommensurate Kekul\'e spiral (IKS) order~\cite{Zhang2019,Bultinck2020kivc,Lian2021,Kwan2021iks,Wagner2021iks}, depending on the precise magnitude of heterostrain in the structure~\cite{Kwan2021iks,Wagner2021iks}. However, the IKS state is time-reversal symmetric~\cite{Kwan2021iks}, and although the K-IVC features local magnetic moments on the moir\'e-scale, it is also time-reversal symmetric upon spatial averaging~\cite{Bultinck2020kivc}. Neither of these states naturally supports an AHE, and are therefore inconsistent with our observations.

Upon breaking $C_{2}$ symmetry, the tBLG bands acquire a staggered sublattice potential mass that opens a gap at the Dirac point, separating the eight flat bands into two distinct groups of four. Each moir\'e subband carries a valley Chern number of either $+1$ or $-1$, with subbands in opposite valleys carrying opposite signs of the valley Chern number owing to their time-reversed relationship~\cite{Zhang2019ahe,Bultinck2020ahe}. Under these conditions, symmetry-broken states with finite valley imbalance and broken time-reversal symmetry can arise more naturally. Both of our devices exhibit a base temperature resistivity above 25 k$\Omega$, nearly an order of magnitude larger than typical tBLG devices~\cite{Cao2018a,Cao2018b,Yankowitz2019,Lu2019}, likely indicating weak $C_{2}$-symmetry breaking from close (but not exact) rotational alignment with the encapsulating boron nitride (see Methods and Supplementary Fig.~\ref{fig:S_CNP} for further discussion). 

\begin{figure}[t]
\includegraphics[width=2.5 in]{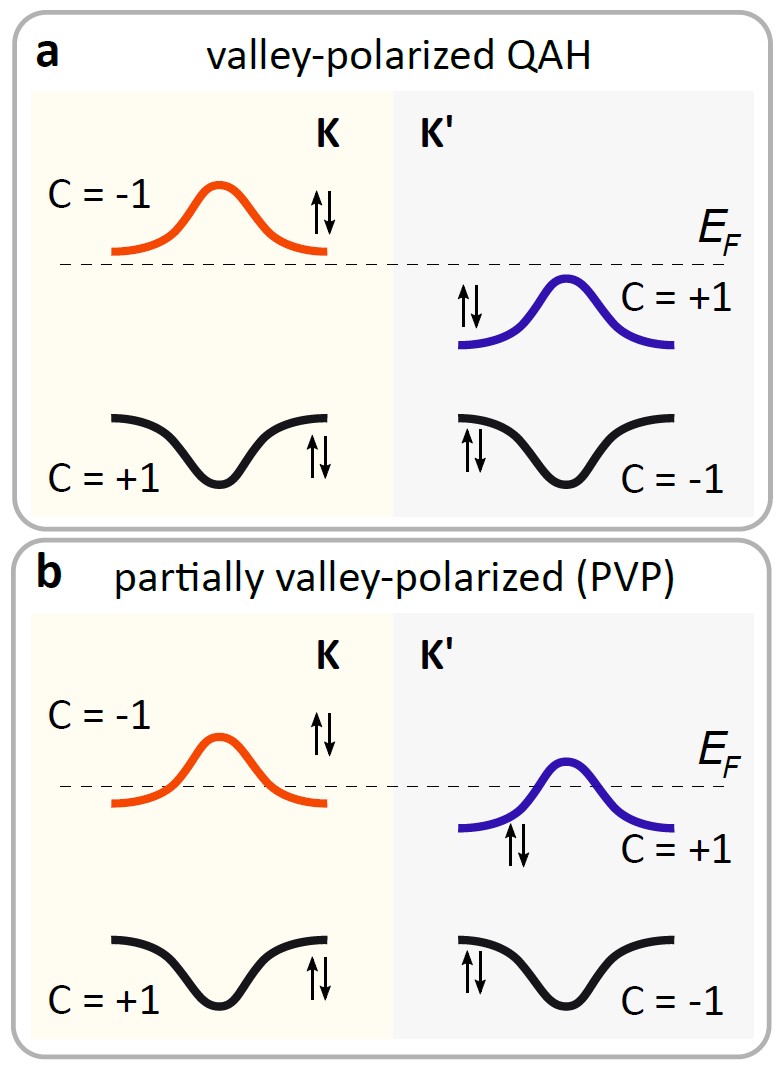} 
\caption{\textbf{Schematic illustration of candidate ground states with AHE at half filling.}
\textbf{a}, The valley-polarized QAH state at $\nu=+2$ has two unoccupied bands from the same valley (depicted as valley K here). The state is gapped with a Chern number of $+2$.
\textbf{b}, The partially valley-polarized (PVP) state is ungapped at half filling, but features an imbalanced occupation of the moir\'e valley subbands. Although there is no quantized Chern number describing the state, there is an overall net Berry curvature that supports a metallic AHE.
}
\label{fig:4}
\end{figure}

At $\nu=-2$ ($+2$), two of the eight moir\'e subbands are (un-)occupied. In this case, there are a handful of nearly degenerate ground states predicted by Hartree-Fock calculations in the absence of $C_2$ symmetry, including a valley-polarized, spin-unpolarized quantum anomalous Hall (VP-QAH) state in which both (un-)occupied bands have the same valley index but are spin unpolarized (illustrated schematically at $\nu=+2$ in Fig.~\ref{fig:4}a), a spin-polarized, valley-unpolarized state in which both (un-)occupied bands have the same spin index but are valley unpolarized, and a valley Hall (VH) state in which the two (un-)occupied bands carry both opposite spin and valley index. Although a disordered and/or small-gap $C=2$ VP-QAH state is consistent with our observation of a metallic AHE, it is energetically disfavored compared with the VH state due to an intervalley Hund's coupling~\cite{Zhang2019,Khalaf2020,Wagner2021iks,Lin2022}. Since only the VP-QAH state supports the AHE from among the viable options at half filling, there appears to be no symmetry-broken ground state in the strong-coupling limit naturally consistent with our observations.

Correlated ground states beyond those considered above have also been identified in Hartree-Fock calculations upon reducing $U/W$ in tBLG with broken $C_2$ symmetry. For instance, Ref.~\cite{Bultinck2020ahe} finds a first-order phase transition to a partially valley-polarized (PVP) state below a critical value of $U/W$, in which all of the isospin subbands are partially occupied but are filled unevenly owing to a Stoner transition (illustrated schematically at $\nu=+2$ in Fig.~\ref{fig:4}b). The PVP state is inherently metallic, but supports a finite orbital magnetic moment due to the spontaneous imbalancing of the valley occupation. However, there are also competing partial VH and partially spin-polarized states that do not support the AHE, and it is not currently clear which ground state should be favored. Understanding our unexpected observation of the AHE at half filling thus requires a more rigorous theoretical treatment of tBLG, very likely in the intermediately-coupled regime.

Finally, we discuss some of the implications of our findings in the context of other recent observations in tBLG. The AHE has been reported at $\nu=+2$ in a magic-angle tBLG device assembled on a monolayer of WSe$_2$~\cite{Lin2022}. However, in this case the proximity-induced SOC in the tBLG couples the spin and valley degrees of freedom, and a generalized Hund's term explicitly favors the VP-QAH state at $\nu=2$. A separate report of a gate-tunable Josephson junction tBLG device (without WSe$_2$) revealed evidence for some form of magnetic ordering at $\nu=-2$~\cite{DiezMerida2021}. Compressibility measurements performed in a large parallel magnetic field in other devices have also raised the possibility of magnetic ordering at half filling~\cite{Park2021hunds}. However, the AHE was not reported in any of these devices, and the usual trivial correlated insulator was observed instead. The origin and nature of the magnetic order underlying the AHE in our devices is therefore very likely to be different from these prior reports in tBLG. Orbital magnetism has been observed at $\nu=-2$ in a magic-angle tBLG device with strong dielectric screening of the Coulomb interactions, and may be of similar nature to that found here, but required the assistance of a perpendicular magnetic field to emerge over a competing zero-field ground state~\cite{Stepanov2020}. Lastly, we note that we do not observe any signatures of superconductivity in our devices (Figs.~\ref{fig:1}a-b), despite it having been seen previously in devices with similar twist angles~\cite{Codecido2019,Saito2020,Cao2021nematic}. Although it is impossible to rule out moir\'e disorder as the cause of its absence, a more likely possibility is that the valley-imbalanced ground state order we observe at half filling is incompatible with pairing. Our results highlight the need for further measurements of devices in the intermediately-coupled regime to help disentangle the potential interplay of various forms of symmetry-breaking and superconductivity.

\section*{Methods}

\textbf{Device fabrication.} tBLG devices were fabricated using the ``cut-and-stack'' method~\cite{Saito2020}, in which exfoliated graphene flakes are isolated using an atomic force microscope tip, and then stacked atop one another at the desired twist angle. Samples were assembled using standard dry-transfer techniques with a polycarbonate (PC)/polydimethyl siloxane (PDMS) stamp~\cite{Wang2013}. Both tBLG devices are encapsulated in flakes of BN with a graphite bottom gate, and then transferred onto a Si/SiO$_2$ wafer. Device D2 additionally has a graphite top gate. The temperature was kept below 180$^{\circ}$C during device fabrication to preserve the intended twist angle. Standard electron beam lithography, CHF$_3$/O$_2$ plasma etching, and metal deposition techniques (Cr/Au) were used to define the complete stack into Hall bar geometry~\cite{Wang2013}. Device D2 was measured while mounted onto a piezo-based apparatus capable of applying uniaxial strain to the device. All results here are acquired without any bias applied to the piezos, which corresponds closely to zero external strain (the results of the strain measurements will be discussed elsewhere).

\textbf{Transport measurements.} Transport measurements below 1.5 K were performed in Bluefors dilution refrigerators with heavy low temperature electronic filtering. Measurements above 1.5 K were performed in a Cryomagnetics VTI or in a PPMS DynaCool. All meaurements were conducted in a four-terminal geometry with a.c. current excitation of 1-10 nA using standard lock-in techniques at a frequency of 17.8 Hz. A voltage was applied to the Si gate in order to dope the region of the graphene contacts overhanging the graphite back gate to a high charge carrier density and reduce the contact resistance. Supplementary Fig.~\ref{fig:S_devices} shows optical microscope images of the two devices. Unless explicitly noted otherwise, all data from Device D1 (D2) was acquired using contacts C-E (A-B) for $\rho_{xx}$ and E-F (A-F) for $\rho_{xy}$.

\textbf{Twist angle determination.} The twist angle, $\theta$, is first determined from the values of the charge doping, $n$, at which the insulating states at full band filling ($\nu=\pm4$) appear, following $n=8\theta^2/\sqrt{3}a^2$, where $a=0.246$~nm is the lattice constant of graphene. It is then confirmed and refined by fitting the quantum oscillations observed at high field (see Fig.~\ref{fig:3}a-c for an example). The filling factor is defined as $\nu=\sqrt{3}\lambda^2 n/2$, where $\lambda$ is the period of the moir\'e. We extract a twist angle inhomogenity of less than $0.05^{\circ}$ in our devices (see Supplementary Fig.~\ref{fig:S_devices}).

\textbf{Absence of the AHE away from half filling.} We do not see evidence of the AHE at any filling factor outside of a small range of doping surrounding $|\nu|=2$. One possibility is that the symmetry-broken states at $|\nu|=1$ and $3$ are inherently non-magnetic in our devices. Another is that symmetry breaking at these integer fillings requires assistance from a magnetic field, thus precluding the AHE at $B=0$. Although we see weak zero-field resistance bumps at certain odd values of $\nu$, none show clear insulating features normally associated with symmetry-broken states in tBLG. A third possibility is that an intrinsic, but weak, AHE at odd $\nu$ could be completely obscured by the twist angle inhomogeneity. We are unable to experimentally distinguish these three cases.

\textbf{Coexistence of competing ground states in Device D2.} We observe sharp dips in the resistivity of Device D2 at filling factors slightly below $\nu=\pm4$, $\pm2$, and $+3$ (Fig.~\ref{fig:1}b). These features project vertically upon applying a magnetic field. The resistance additionally becomes negative as $B$ is increased near $\nu=\pm4$ and $-2$, shown in white in Fig.~\ref{fig:1}b. Four-terminal measurements of an insulating state often result in negative-resistance artifacts, especially in the presence of inhomogeneity, as the potential difference between the voltage probes can invert in the percolative transport regime. These artifacts therefore reveal the coexistence of the trivial insulators typically observed at these band fillings in magic-angle tBLG, presumably originating from a portion of the device with slightly smaller twist angle ($\theta \approx 1.15^{\circ}$). Supplementary Fig.~\ref{fig:S_D2} shows additional transport measurements acquired with other contacts on the sample. The coexisting trivial correlated insulator at $\nu=-2$ is most clear in the contacts at the top of the device (corresponding to the region with AHE), and becomes progressively weaker in contacts towards bottom of the device where the measured twist angle becomes slightly larger. The coexistence of the trivial correlated insulator and AHE states at $\nu=-2$ implies that they are likely nearly degenerate in this sample, with small changes in the twist angle responsible for the switching between the two. We note that we do not observe any similar features in Device D1 (see Supplementary Fig.~\ref{fig:S_D1} for measurements of additional contacts).

\textbf{Potential alignment with the encapsulating boron nitride.} It is highly likely that the magnetic ground state we observe at half filling requires $C_2$-symmetry breaking in order to support a finite valley polarization (although there could be more exotic alternatives we have not considered here). In principle, $C_2$-symmetry can break spontaneously, however it is typically observed when the tBLG is within a few degrees of rotational alignment with one of the encapsulating boron nitride (BN) crystals. A large AHE approaching quantization to $h/e^2$ was observed at $\nu=+3$ in previous tBLG devices with BN misalignment of less than $\sim1^{\circ}$, most simply understood as arising from a complete spin and valley polarization~\cite{Sharpe2019,Serlin2020}. Our observations contrast these results, however, as we instead observe a metallic AHE only at $\nu = +2$ or $-2$. 

Although we do not have a direct measure of the twist angle of either encapsulating BN crystal with the tBLG, we do not observe any signatures of additional superlattice minibands indicative of near-$0^{\circ}$ alignment. Nevertheless, we do see an unusually large CNP resistivity in our devices (Figs.~\ref{fig:1}a-b and Supplementary Fig.~\ref{fig:S_CNP}), as well as a weak thermal activation behavior in Device D1 (we did not measure this carefully in Device D2). Previous measurements of BN-encapsulated monolayer graphene have revealed the formation of meV-scale gaps at the graphene CNP even in devices in which the rotational misalignment of the BN approaches $5^{\circ}$~\cite{Hunt2013} --- well beyond the ability to reach full filling of the resulting moir\'e miniband with gating. We therefore speculate that both of our devices are likely somewhat (but not precisely) aligned to at least one of the encapsulating BN crystals, providing a natural mechanism for $C_2$-symmetry breaking. The orbital magnetism may depend sensitively on the precise alignment and near commensurabilities of the twisted graphene and graphene/BN moir\'e superlattices~\cite{Shi2021}, and the local fluctuations of the atomic registries may provide an additional source of magnetic disorder in the form of a rapid spatial fluctuation in the sign of the local mass gap~\cite{Grover2022}.

\section*{Data Availability}
Source data are available for this paper. All other data that support the plots within this paper and other findings of this study are available from the corresponding author upon reasonable request.

\section*{Acknowledgments}
We thank Yahui Zhang, Nick Bultinck, and Zhi-Da Song for helpful discussions. This work was supported as part of Programmable Quantum Materials, an Energy Frontier Research Center funded by the U.S. Department of Energy (DOE), Office of Science, Basic Energy Sciences (BES), under award DE-SC0019443; the Army Research Office under Grant Number W911NF-20-1-0211 to M.Y.; and the Gordon and Betty Moore Foundation’s EPiQS Initiative, grant no. GBMF6759 to J.-H.C. J.-H.C. also acknowledges support from the David and Lucile Packard Foundation. M.Y. and J.-H.C. acknowledge support from the State of Washington funded Clean Energy Institute. This research acknowledges the usage of a dilution refrigerator system that was provided by NSF DMR-1725221, and of the millikelvin optoelectronic quantum material laboratory supported by the M. J. Murdock Charitable Trust. K.W. and T.T. acknowledge support from the Elemental Strategy Initiative conducted by the MEXT, Japan (Grant Number JPMXP0112101001) and JSPS KAKENHI (Grant Numbers 19H05790, 20H00354 and 21H05233).\\

\section*{Author contributions}
C.-C.T. and X.M. fabricated the devices. C.-C.T., X.M. and Z.L. performed the measurements. K.W. and T.T. provided the bulk BN crystals. C.-C.T., X.M., Z.L., J.-H.C. and M.Y analyzed the data and wrote the paper.

\section*{Competing interests}
The authors declare no competing interests.

\section*{Additional Information}
Correspondence and requests for materials should be addressed to M.Y.

\section*{Supplementary Information}
Supplementary Figs.~S1-S6.

\bibliographystyle{naturemag}
\bibliography{references}

\clearpage

\renewcommand{\thefigure}{S\arabic{figure}}
\renewcommand{\thesection}{S\arabic{section}}
\renewcommand{\thesubsection}{S\arabic{subsection}}
\renewcommand{\theequation}{S\arabic{equation}}
\renewcommand{\thetable}{S\arabic{table}}
\setcounter{figure}{0} 
\setcounter{equation}{0}
\appendix 

\onecolumngrid

\section*{Supplementary Information}

\begin{figure*}[h]
\centering
\includegraphics[width=6.9 in]{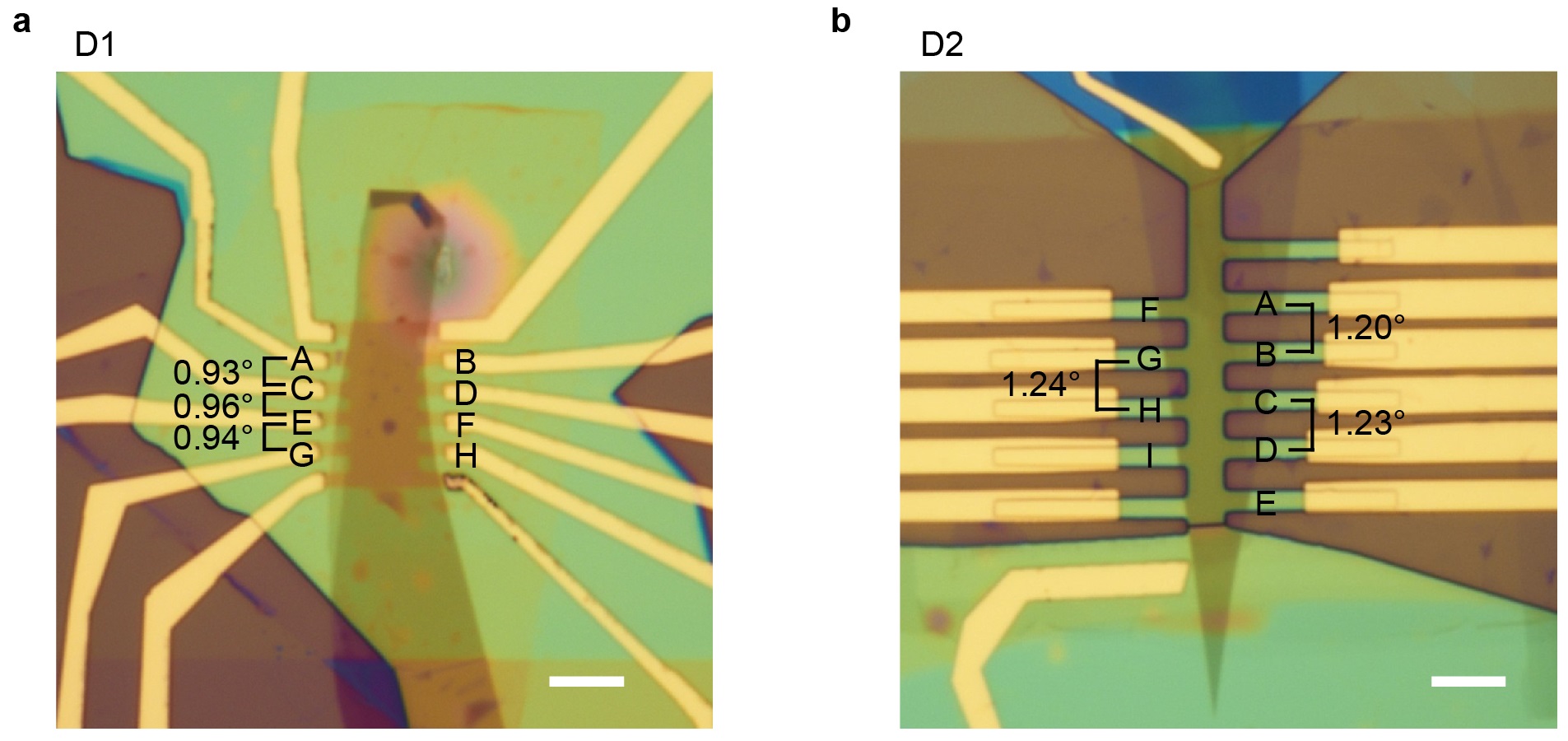} 
\caption{\textbf{Optical microscope images of the devices.}
\textbf{a}, Device D1, and \textbf{b}, Device D2. The twist angle measured between neighboring pairs of contacts is also shown. The scale bars are 5~$\mu$m.
}
\label{fig:S_devices}
\end{figure*}

\begin{figure*}[h]
\centering
\includegraphics[width=6.9 in]{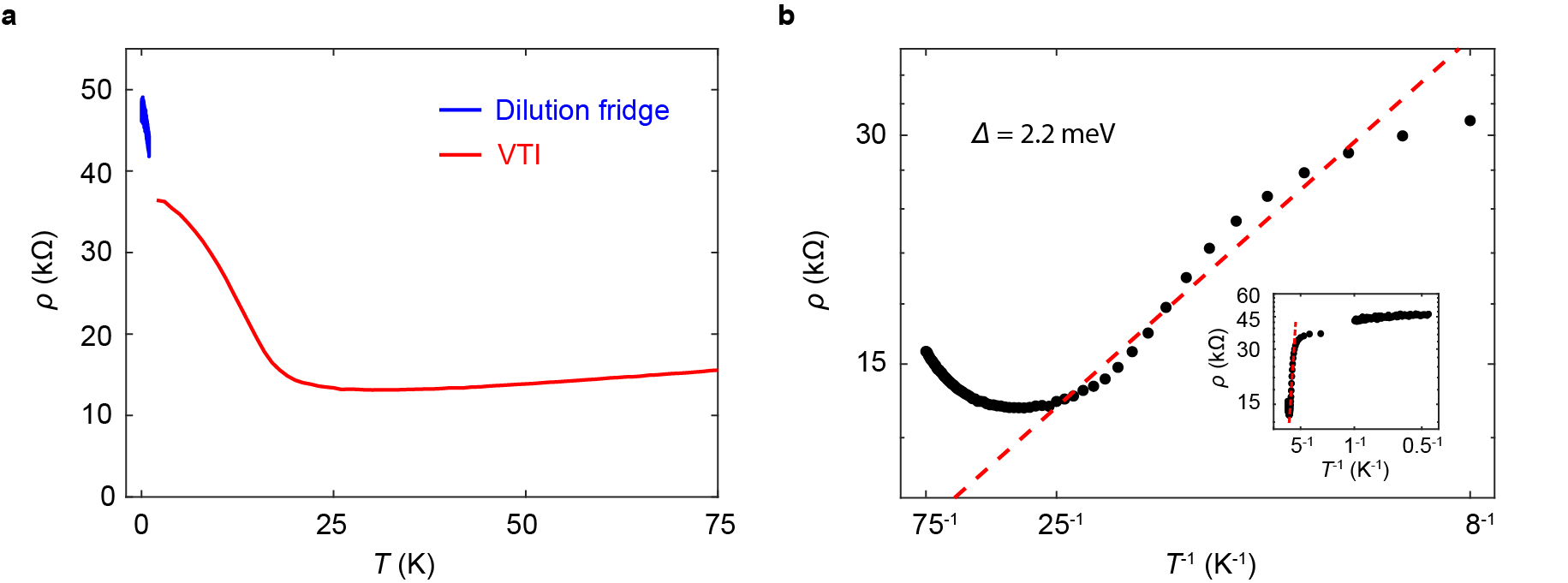} 
\caption{\textbf{Thermal activation of the CNP in Device D1.}
\textbf{a} Resistivity of Device D1 at the CNP ($\nu=0$) measured as a function of temperature, exhibiting insulating behavior below $\sim25$~K. The red curve is measured in a VTI down to 1.5 K, and the blue curve is measured in a dilution fridge down to 20 mK.
\textbf{b} The same data shown on an Arrhenius plot. The CNP exhibits a small region of (approximately) activated behavior. We extract the band gap, $\Delta=2.2$~meV, from slope of the linear fit (red dashed line) using $\rho \propto e^{\frac{\Delta}{2 k T}}$, where $k$ is the Boltzmann constant.
}
\label{fig:S_CNP}
\end{figure*}

\begin{figure*}[h]
\centering
\includegraphics[width=6.9 in]{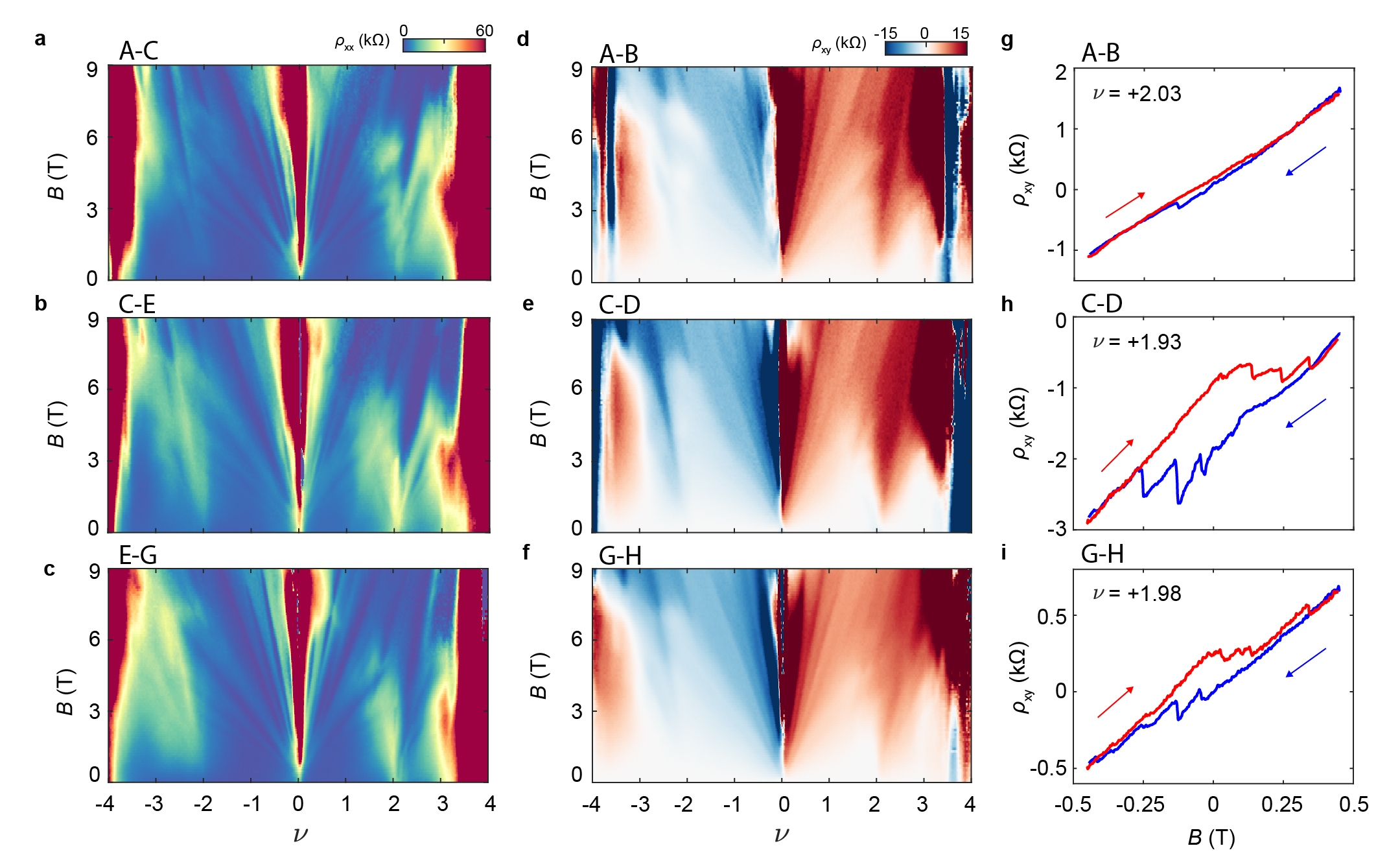} 
\caption{\textbf{Transport measurements from additional contact pairs in Device D1.}
Landau fan diagrams of \textbf{a-c}, $\rho_{xx}$ and \textbf{d-f}, $\rho_{xy}$ measured between different pairs of contacts corresponding to the contact labeling scheme in Supplementary Fig.~\ref{fig:S_devices}a. 
\textbf{g-i}, Measurements of the AHE near $\nu=+2$ taken using the same contacts as the associated $\rho_{xy}$ fans shown in (\textbf{d-f}). The data are acquired at the same gate voltage, which corresponds to a slightly different value of $\nu$ due to the twist angle disorder. All measurements are acquired at $T=20$~mK.
}
\label{fig:S_D1}
\end{figure*}

\begin{figure*}[h]
\centering
\includegraphics[width=6.9 in]{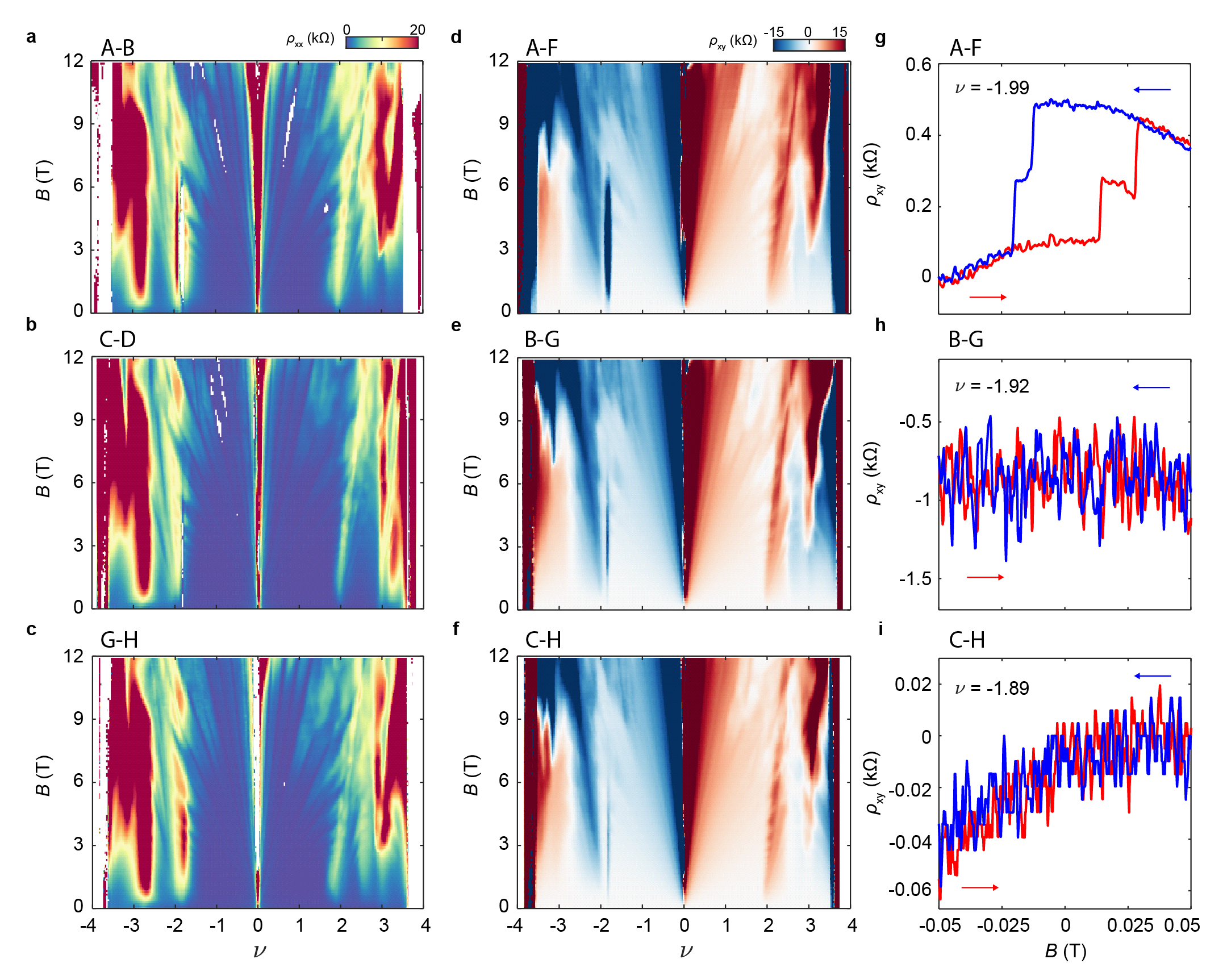} 
\caption{\textbf{Transport measurements from additional contact pairs in Device D2.}
Landau fan diagrams of \textbf{a-c}, $\rho_{xx}$ and \textbf{d-f}, $\rho_{xy}$ measured between different pairs of contacts corresponding to the contact labeling scheme in Supplementary Fig.~\ref{fig:S_devices}b. The strength of the coexisting trivial insulating state at $\nu=-2$ varies substantially depending on the contact pair.
\textbf{g-i}, Measurements of the AHE near $\nu=-2$ taken using the same contacts as the associated $\rho_{xy}$ fans shown in \textbf{d-f}. The data are acquired at the same gate voltage, which corresponds to a slightly different value of $\nu$ due to the twist angle disorder. The AHE is only observed in contact pair A-F, despite the overall similarities of the Landau fans. For contacts B-G in particular, there is a large offset from $\rho_{xy}=0$ due to mixing with $\rho_{xx}$. All measurements are acquired at $T=100$~mK.
}
\label{fig:S_D2}
\end{figure*}

\begin{figure*}[h]
\centering
\includegraphics[width=5 in]{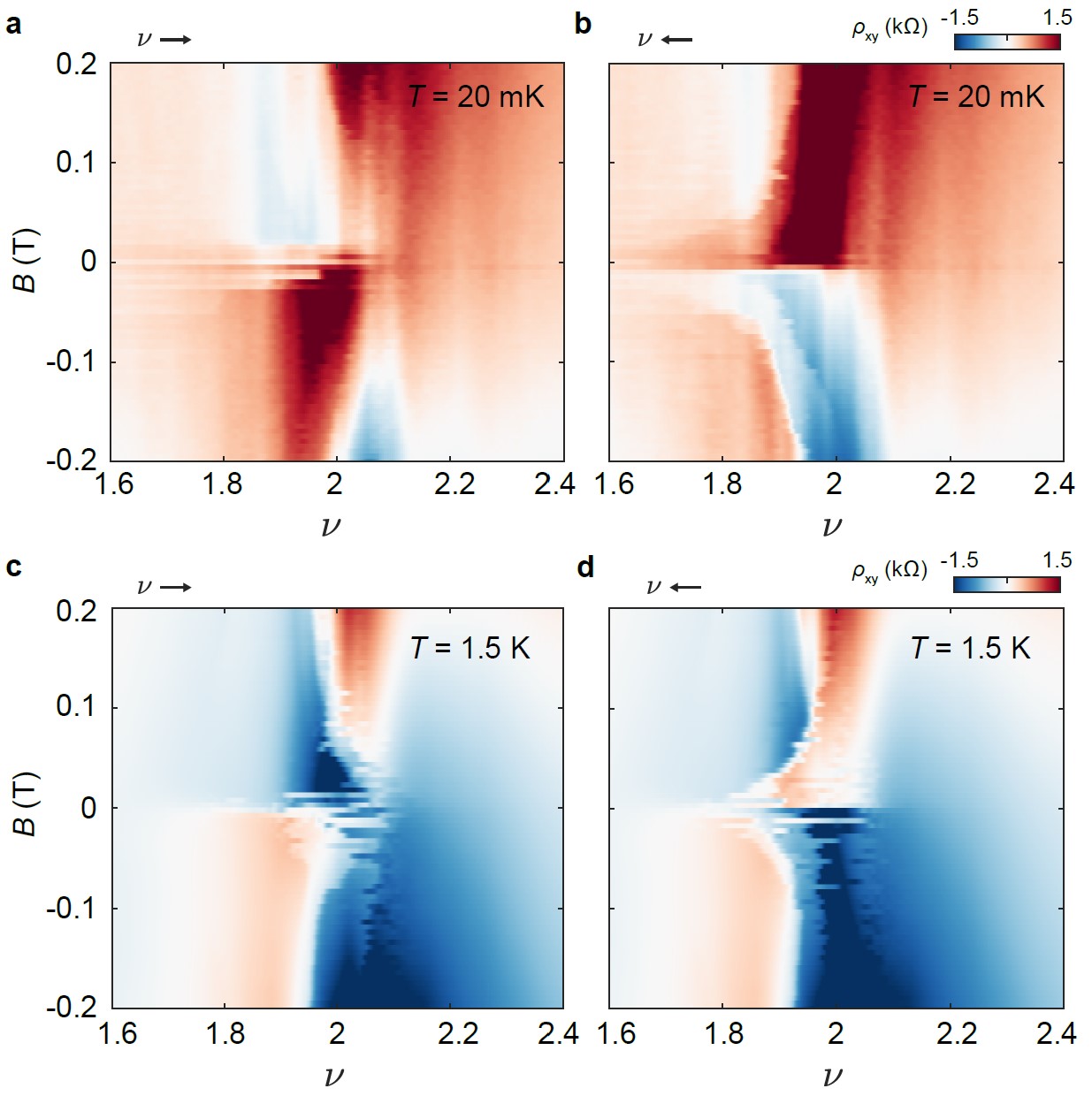} 
\caption{\textbf{Electric-field--reversal of the magnetic state in Device D1.}
$\rho_{xy}$ acquired by sweeping $\nu$ from \textbf{a}, small to large values, and \textbf{b}, large to small values, at fixed $B$ and $T=20$~mK.
\textbf{c-d}, The same measurement at $T=1.5$~K. At both temperatures, we observe hysteresis at low magnetic fields depending on the direction the doping is swept.
}
\label{fig:S_dopingAHE}
\end{figure*}

\begin{figure*}[h]
\centering
\includegraphics[width=4 in]{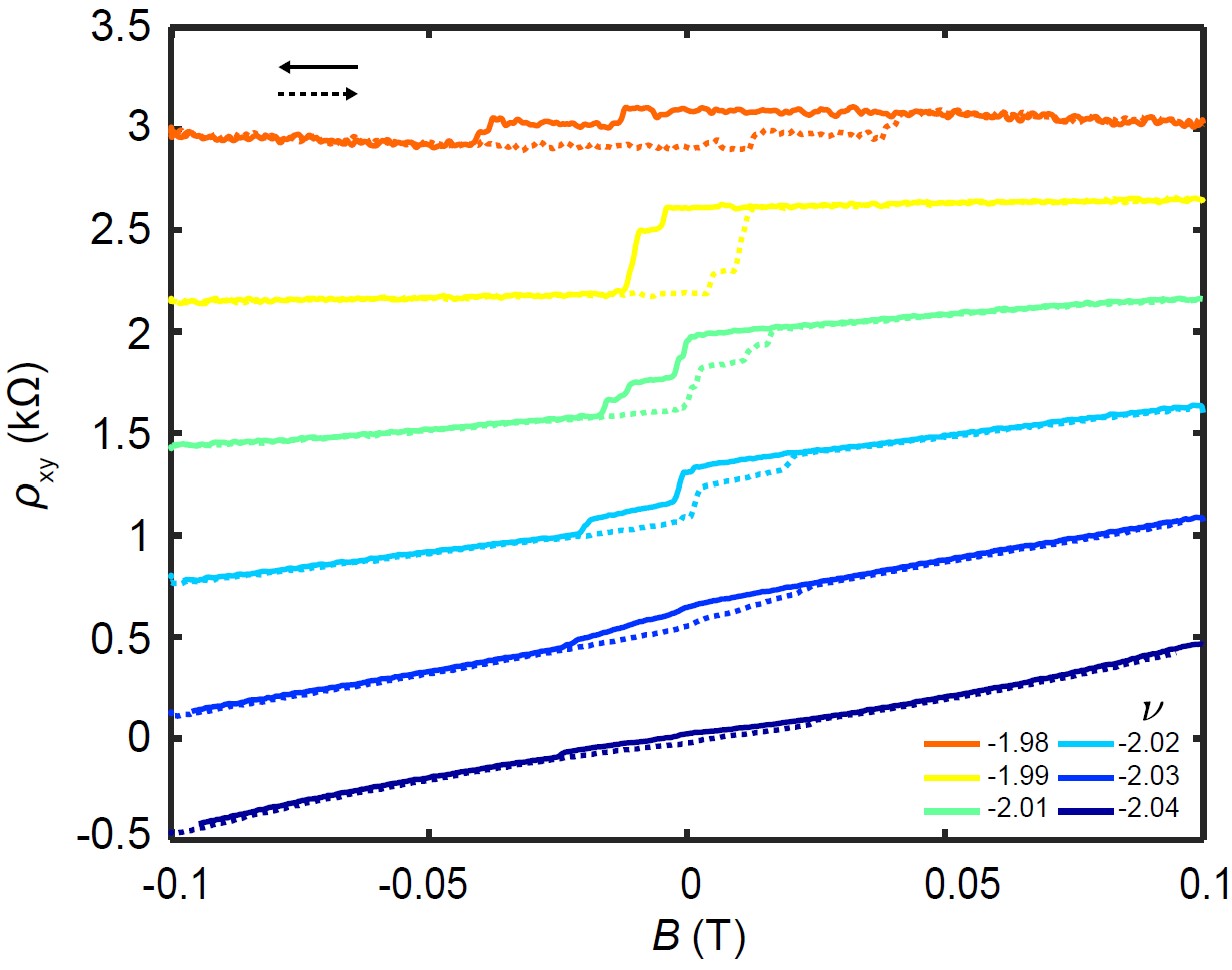} 
\caption{\textbf{AHE versus doping in Device D2.}
$\rho_{xy}$ measured as $B$ is swept back and forth around at selected values of $\nu$ around $\nu=-2$ in Device D2. The measurements are acquired at $T=100$~mK.
}
\label{fig:S_AHE_D2}
\end{figure*}

\end{document}